# So You Want to be a Professional Astronomer!

*Exotic workplace locales, amazing discoveries, and fame (but probably not fortune) await those who persevere on the path leading to a career as a professional astronomer.*

by Duncan Forbes

*Wanted: Astronomer. Must be willing to work occasional nights on the top of a mountain in an exotic location. A sense of adventure and nomadic lifestyle is a plus. Flexible hours and casual dress code compensate for uncertain long-term career prospects and average pay. The opportunity for real scientific discovery awaits the right candidate. Apply now.*

In many ways, professional astronomers are very fortunate. They have an opportunity to continue their passion (one that many people share) *and* they're paid to do it. Some of the reasons given by PhD students for becoming an astronomer include: it's fun and exciting, there are great opportunities for travel, it's a cool job, and it's possible to make significant discoveries.

Universities, observatories, government organizations, and industry employ astronomers who, contrary to popular belief, don't spend all their waking hours at a telescope. Instead, most of their time is spent teaching, managing projects, providing support services, and doing administrative duties. A typical astronomer might spend just a week or two a year on an observing run, followed by months of data analysis and article writing.

If you're going after a career in astronomy, be warned: It is extremely competitive! There are many very smart, hard working people seeking a limited number of positions. The worldwide community of professional astronomers is only about 10,000; most are located in the US (with about 1,000 in the UK and 250 in Australia).

Under the heading of "astronomy" there are many fields (and sub-fields) of research, and if you choose one with few researchers, the conferences you attend will have the feel of a family reunion. (If research isn't your thing, there are non-research options, including Support Astronomer and Telescope Operator, which will let you spend a lot of time around telescopes all over the world.) The American Astronomical Society (AAS) has a useful guide [www.aas.org/education/publications/careerbrochure.pdf] that describes various careers in astronomy.

So how do you join the elite ranks of professional astronomy? Here are some suggestions for how to get, and how *not* to get, a job in astronomy.

**First, Get that PhD!**
All professional research astronomers have a PhD in astronomy or a related field. Use the web and talk to people about the best places to do your PhD. Be bold and choose a different university for your PhD than your undergraduate degree. This exposes you to different ideas and broadens your horizons. It also looks better to a potential employer. You may even consider doing your PhD overseas. Advantages could include a shorter program (three to four years in the UK and Australia vs. five to six years in the US) and no Graduate Record Examination (GRE) required.

Attributes of a good PhD student include a passion for research, a high level of motivation, well organised, and good verbal and written skills. As a student you will probably be working more than 40 hours a week (think apprentice), so it's important to work efficiently. The old mantra "work smarter, not harder" is very relevant here, especially as data volumes continue to grow at an exponential rate. Two good articles on what it's like to be a PhD student and how to obtain a PhD are:
"How to be a Good Graduate Student"
[www.cs.indiana.edu/HTMLit/how.2b/how.2b.html] and "So Long, and Thanks for the PhD!" [www.cs.unc.edu/~azuma/hitch4.html].

Choose your PhD supervisor carefully. They will be your guide and mentor for the next few years. It's a good idea to check out their publication record to see where their recent interests lie, and ask current students what they think of their supervisor and the research group/department. There is a wide range of supervisory styles from the "Hi, here is a research topic. Come back and see me in 3 years time." to "I want updates of your progress every 5 minutes."

Some supervisors can be quite demanding, which likely stems from two factors — their research reputation is at stake, too, and they want to prepare you for the 'real world' of independent research. Richard Reis has written several interesting articles in the Chronicle for Higher Education which includes "Choosing the Right Research Advisor" [http://chronicle.com/jobs/2000/01/2000010702c.htm].

While working on your PhD, you should aim to write papers (and publish them!) as you go. This will make the actual writing of your thesis a much easier task. I suggest you try for one published paper for each year of full-time research. Some students manage more than a half dozen papers during their PhD program. The bad news is that you have to compete with them in the job market! And don't forget to read other people's papers, because 'knowing the literature' is very important.

It's also a good idea to discover the 'big picture' beyond your narrow sub-field, I suggest spending about 10% of your week attending seminars and chatting with colleagues outside your field about their work. Some collaboration work done outside of your department will look good when it comes time for letters of reference and job applications.

Networking is also important for your career so hone your skills during your time as a PhD candidate. Give research talks. Being able to present your research can be crucial to your career prospects, so get plenty of practice. Finally, consider applying for small grants and awards as these can help improve your CV.

Warning: Too much time spent observing or writing computer code can adversely affect your chances of acquiring a PhD! While this work might form the basis of your project, be careful it doesn't become all-consuming – you still need to prepare and present a thesis to be awarded a PhD.

**Becoming a Postdoc**
At some point toward the end of your PhD work, it'll be time to apply for a post-doctoral research position. The best place to look for a postdoc (or staff) position at a university or observatory is the monthly AAS Job Register [http://members.aas.org/JobReg/Jobregister.cfm]. Each year some 200 short-term postdoc (and about 80 permanent staff) positions are advertised worldwide, with peak activity occurring in November.

Postdocs can be divided into 'named' and 'unnamed' positions. The named positions include Hubble and Chandra Fellowships in the US, and Fellowships funded by the national research councils in the UK and Australia. These positions generally offer freedom to explore your own research direction, a (reasonably) generous research budget, and a decent salary. As such they are prestigious and highly competitive. Unnamed positions are typically with individual astronomers or university departments that have generated funds for the position via a research grant, and the research topic is likely predetermined.

In either case, you may be invited to join a large team. Being a member of a large research group can allow you to tackle major scientific questions and work with top people in your field. However, it can also make it difficult for people outside the team to evaluate your contribution to the project.

**First, the good news!** Although most countries overproduce astronomy PhDs relative to their job market, the number of postdoc positions worldwide roughly matches the demand for positions (after excluding people who don't wish to continue in astronomy or are unwilling to live abroad). In the most recent decadal report of Australian astronomy, some 70% of PhD recipients obtained a postdoc (mostly abroad), 20% obtained a job in industry, and 10% don't respond to questionnaires.

So generally speaking there *is* a postdoc position in astronomy for you if you want it. Postdocs are the key period in which you show what you're made of in terms of the quality and quantity of your publications. The average academic astronomer in the UK produces 4.4 papers a year. Ambitious young postdocs should be looking to match or exceed that level with quality papers. A typical research career involves two to three postdocs each lasting two to three years. The next step is an application for an entry-level Lectureship or an Assistant Professor job.

**Now the bad news.** It's tough to get a *permanent* job in astronomy. It is not unheard of for a university department to receive more than 100 applications for a single position. Although the numbers vary over the years, a recent report by the UK's Royal Astronomical Society concluded that only 1 in 5 students with a PhD in astronomy obtained a permanent job in the field in the medium term — meaning by the time the "student" is about 40 years old!

It's also worth bearing in mind that the popularity of sub-fields in astronomy, and hence the number of related jobs that are available, change with time. In a survey of Australian astronomers (covering the period 1995 to 2000), the percentage who said that they were working in galactic astronomy declined from 41% to 24%, while the fraction of those exploring extragalactic topics rose from 26% to 42%.

**Movin' on up**
If you want to move up the job ladder, you'll have to evolve from an apprentice-like PhD student to a research leader or manager. You will find yourself making smaller contributions to more papers. You'll have a better grasp of the big picture but probably at the expense of the technical details. Choosing your collaborators well is an important aspect of ongoing research success. You will increasingly multi-task, juggling teaching, community service, administration, management, personnel, and finance issues along with your research and that of your students.

Your first step in this evolution is to leave the world of the postdoc and acquire that permanent position. You'll need to apply, of course, and the better your CV, the better your chances. Your written application (including cover letter, CV, research interests, and letters of reference) is key to getting a job interview. Give considerable thought as to whom to ask to write those reference letters. It's obviously good if the writers are well regarded by your potential employer, but it is equally important to get a strong letter from someone who knows you well.

When it comes time to apply for a job, you'll likely be inundated with advice and suggestions. So let me suggest what you **shouldn't** do.
- Use the 'shotgun' approach of applications: many and wide.
- Don't read the application instructions.
- Write it on the last possible day.
- Fail to run the spellchecker.
- Fail to include a well-directed cover letter.
- Don't get a senior colleague to read your application.
- Don't tell your referees you have put their names forward.
- Or tell them, but not until the day before the deadline.

When you get a job interview, be prepared and do your homework. Think about *why* you want the job — it's probably the first question you'll be asked. You may also be asked potentially tricky questions like: "What are your career plans?" and "If offered this job today, would you accept it?" It's also a good idea to have some questions of your own lined up. There are plenty of websites and books with strategies on how to interview well — look at a few beforehand.

Speaking of the web, astronomy job webpages with the latest rumors and gossip about positions (and what it's like to work at various institutions) have added an interesting new dimension to the application and hiring process. On the flipside, an employer may Google you. So you might consider cleaning up your personal web page, including any public *MySpace* or *Facebook* entires.

If you're invited to visit your potential employer, you may be asked to give a seminar on your research. This will form a crucial part of your job interview, but how *not* to give a research talk is a topic for another time.

**Publications: Quantity *and* Quality**
Once you acquire that coveted permanent position, your life will revolve around teaching, doing research, and publishing your research. Why do we publish? As scientists we need to communicate the results of our research, a published paper is our

'product,' and (like it or not) these papers are a measure of our productivity. *Not publishing your results will result in a remarkably short astronomical career.*

The number of papers posted to the astro-ph preprint server [http://lanl.arxiv.org/archive/astro-ph] has risen steadily since 1992, and this increase shows no sign of abating. One reason for astro-ph's popularity is that if you publish only in a journal and do *not* post your paper online, you may decrease its citation rate by half.

In 2007, the number of papers posted to astro-ph exceeded 10,000. *This translates to more than 40 new papers each working day!* Even if you select only the papers in your sub-field, it's still very difficult to keep up. Some astronomers don't even try to.

With so many papers appearing daily, how do you make other astronomers aware of *your* work and get them to cite it? One solution is to tell them what you do by presenting seminars in their departments and by giving talks/posters at conferences. You should also give careful thought to the words contained in the paper's abstract so your paper is easy to find by someone doing an abstract-based search.

While many funding agencies and employers look only at the *quantity* of your papers, the *quality* of your publications is arguably a much more important measure. Quality in this context is often taken as the impact of your publication on other astronomers and for that we use the number of citations to your paper.

Although Scopus [www.scopus.com] and Thomson Scientific [http://scientific.Thomson.com] track citations, the most up to date source for astronomers is the Astrophysical Data Service (ADS) [http://adsabs.harvard.edu/abstract_service.html], which gives both raw citations and citations normalised by the number of authors. In 2004 Frazer Pearce compiled the relative distribution of all raw and normalised ADS citations for astronomers ("Citation measures and impact within astronomy") [http://lanl.arxiv.org/abs/astro-ph/0401507] and found that the top 10% of active astronomers worldwide typically have 382 raw and 74 normalised citations in the previous five years.

**Just Do It**
In summary, the three steps to job success in astronomy (with apologies to Nike) are:
1) Research it.
2) Publish it.
3) Talk about it.

Repeat steps 1 to 3 several times a year, and a long career in professional astronomy awaits you. In the process, if you discover something significant and become famous, then so much the better. Don't forget to network, always keep the big picture in mind, and enjoy yourself.

This article is based on discussions with PhD students, postdoctoral researchers, and senior colleagues I have worked with over the years, particularly in the US, UK, and Australia. I hope it sheds some light on the process of landing a job in astronomy and is useful to anyone seeking a long-term research career in astronomy.

Duncan Forbes *certainly made some mistakes when in the astronomy job market, but he survived and is now a Professor at Swinburne University in Australia, after a Lectureship in the UK and a postdoc position in the US. His research interests include globular clusters and galaxy formation. He thanks everyone who has contributed to the discussions that helped crystallize this article. He particularly thanks Anna Russell, Alister Graham, Frazer Pearce, and Jay Strader for their input. This article was written under cloudy skies at Siding Spring Observatory in Australia.*